# Dynamics characterization of the glass formation of twist-bend liquid crystal dimers through dielectric studies

M. Czarnecka[1], Y. Arakawa[2], A. Kocot[3] and K. Merkel[3*]

[1] Faculty of Electrical Engineering, Automatics, Computer Science and Biomedical Engineering, AGH University of Science and Technology, al. Adama Mickiewicza 30, Cracow, 30-059 Poland

[2] Department of Applied Chemistry and Life Science, Graduate School of Engineering, Toyohashi University of Technology, Toyohashi 441-8580, Japan

[3] Institute of Materials Engineering, Faculty of Science and Technology, University of Silesia, 75 Pułku Piechoty 1a, Chorzów 41-500, Poland

**ABSTRACT**

Broadband dielectric spectroscopy in a frequency range of 10 to $2\times 10^9$ Hz were used to study the molecular orientational dynamics of the glass-forming, thioether-linked cyanobiphenyl liquid crystal dimers CBS7SCB and CBS7OCB. As was expected theoretically, two different relaxation processes that contributed to the dielectric permittivity of dimers were observed. The low-frequency relaxation mode, m1, was attributed to an "end-over-end rotation" of the dipolar groups parallel to the director. The high-frequency relaxation mode, m2, was associated with the precessional motions of the dipolar groups around the director. The relaxation times for both modes were analyzed over a wide temperature range down to near the glass transition temperature. The different analytic functions that were used to characterize the temperature dependence of the relaxation times of the two modes are discussed. Particularly, the critical-like description *via* the dynamic scaling model seemed to give not only quite good numerical fittings, but also provided a consistent physical picture of the orientational dynamics on approaching the glass transition. When compared to the IR spectroscopy finding, in the higher temperature region of $N_{TB}$ phase, where the longitudinal correlations of dipoles grew, the m1 mode experienced a sudden increase of enthalpy while m2 changed continuously, which is described by the critical mode coupling behavior. Both types of molecular motion appear to strongly cooperate at a low temperature range of the $N_{TB}$ phase, but changed in a coordinated manner as the temperature of the material approached the glass transition point. As was expected, it was found that both molecular motions determined the glass dynamics at the same glass transition temperature, Tg.

**INTRODUCTION**

The simplest liquid crystalline mesophase, the uniaxial nematic (N) phase, is characterized by an orientation of the molecules in a preferred direction, but with a low positional order. The introduction of chirality, through either a chiral center in the mesogen or by adding a chiral dopant to the N phase of the host, results in the formation of a chiral nematic (N*) phase that has a helical structure. Independently, Meyer [1] and Dozov [2] predicted that liquid crystal phases with a local chirality could be formed by bent achiral mesogens. The helical phase could occur spontaneously as a result of the simultaneous bending and twisting of a local director in an array of non-chiral molecules. The resulting twist-bend nematic ($N_{TB}$) phase is present in an equal number of degenerate domains that have opposite twist directions and the director is tilted relative to the helical axis. This new nematic-type phase is distinguished by unusual periodic patterns that can be observed using polarizing microscopy and a very rapid electro-optic switching in the microsecond regime [3,4]. However, an unambiguous determination of the phase structure appears to be very difficult as it exhibits no modulation of the electron density. Extensive studies [5-7] have suggested that this structure corresponds to a phase with a spontaneous conical twist-bend director distortion ($N_{TB}$), which was theoretically predicted to be driven by an elastic instability with a sign inversion of the bend elastic constant $K_{33}$ for bent-shaped molecules [2].

The $N_{TB}$ phase was first discovered for 1,7-bis-4-(40 -cyanobiphenyl)heptane, CB7CB, which consists of two mesogenic units that are connected by a flexible spacer [5-8]. The requisite bent molecular shape is exhibited in dimers with a hydrocarbon spacer of an odd parity. Other dimeric mesogen [9-17] bent-core species [18-20] have been reported to manifest the $N_{TB}$ phase. Despite the significant number of materials that have been investigated, a general structure - property relationship for the $N_{TB}$ phase remains elusive; although it has been found that a spatially uniform curvature of the molecule is necessary to form the $N_{TB}$ phase regardless of the underlying chemical groups [21,22].

The primary motivation for this work was to analyze the orientational dynamics of the molecules in the nematic phases. N and $N_{TB}$, which are formed by the polar symmetric dimers CBSCnSCB and the similar asymmetric dimers CBSCnOCB. Dielectric relaxation measurements were performed for both samples in the entire temperature range up to the glass transition. The observed relaxation processes, like those in earlier dielectric relaxation studies [5,20,23-24], were interpreted using the different molecular motions that are defined by the theoretical models of dielectric relaxation [25-27]. In both nematic phases, the rotation of the dipolar groups that are associated with the terminal cyanobiphenyl groups led to two relaxation modes that were related to the rotational dynamics of the molecules. The low-frequency mode, denoted by m1, was caused by an "end-to-end" motion of the dipolar groups, which were parallel to the mesogen axis. The high-frequency mode, denoted m2, was the result of the precessional rotation of the dipolar groups around the director. These modes contributed to the complex dielectric permittivity differently depending on the orientation of the dielectric relative to the measured electric field.

An important part of the analysis concerned the study of the glass transition, which was clearly identified in both of the longer dimers below the N-$N_{TB}$ transition. The glass transition was characterized by the temperature $T_g$ and the rate at which the various properties of the material changed with the temperature as the liquid phase approached the glass transition. For systems that exhibit a certain number of structural relaxations, it is possible to identify the dielectric glass transitions that correspond to each of the two types of molecular motion independently. It is interesting whether changes in the molecular interactions that were observed in previous spectroscopic studies [28,29] might be related to the observed glass transitions. The results for the CBSC7SCB and CBSC7OCB dimers presented here are discussed for the information on which molecular movements are frozen during glass formation. The results for the symmetrical and asymmetric liquid crystal dimers that are presented here are discussed in order to provide information on how the changes in the molecular interactions that are observed near and below the N-$N_{TB}$ transition affect the aggregation process and how they are frozen during glass formation.

**RESULTS AND DISCUSSIONS**

**1. Dielectric relaxation**

The simplest dielectric absorption model that is applicable to most materials is the Debye model, which describes the single dipole relaxation process. The temperature dependence of the relaxation time can be represented by the phenomenological Arrhenius equation, which introduces the activation energy $E_a$ for the reorientation of a molecular dipole in a dielectric environment.

$$\tau = \tau_0 \exp[\, E_a/k_B T] \qquad (1)$$

Although the Arrhenius equation describes the dielectric relaxation of many simple fluids well, there are several materials for which the equation fails, and it is then necessary to modify the equation in order to describe the experimental results. One of the phenomenological equations that is most commonly used to describe the temperature dependence of the relaxation time data ($\tau$) is the Vogel-Fulcher-Tammann (VFT) equation [30]:

$$\tau = \tau_0 \exp[\, B/(T - T_0)], \qquad (2)$$

Despite some concerns about the validity of the VFT-equation as it assumes a dynamic divergence of the relaxation time at some finite temperature $T_0$, some theories such as the Adam-Gibbs entropy model (AG) [31] and some more recent theoretical models [32-34] are based on the VFT equation. The basic idea behind these theories is that glass formation is associated with highly cooperative movements in a structurally fluctuating sample, whose cooperativity grows as $T_g$ is approached. The dynamic scaling (DS) model [35, 36], which combines the above-mentioned ideas of glass formation with a mean-field description of the virtual phase transition has been proposed such that:

$$\tau = \tau_0 \,[(T - T_c)/T_c]^{-\Phi} \qquad (3)$$

where the pre-factor $\tau_0$ is defined as the relaxation time at $2T_C$; the temperature $T_C$ is the temperature of the virtual phase transition (the critical temperature), which is usually located slightly below $T_g$. For the high-temperature dynamic domain, which is well above the glass transition where the coupling mechanisms can be disregarded, the Mode Coupling Theory (MCT) provides a power law function that is similar to the above eq. but with fitting parameters of the different physical meanings to those that have been reported for the DS-model. In that case, $T_C$ accounts for the crossover temperature from the ergodic to the non-ergodic domain. The critical temperature $T_C$ seems to correlate with the caging temperature $T_A$ [37].

An alternative non-divergence description that has recently been considered is the Waterton-Mauro (WM) parameterization, which was derived empirically by Waterton [38] in the 1930s, and was recently derived theoretically by Mauro and colleagues [39] and is now considered to be a promising function for representing relaxation times, which are defined as follows:

$$\tau = \tau_0 \exp[K/T \exp(C/T)], \qquad (4)$$

where $\tau_0$ has the same meaning as in the VFT equation; the $K$ and $C$ parameters are related to the effective activation barriers, which are defined as both of the thermal activation fitting parameters.

The problem is compounded in dielectric studies of liquid crystals because of their macroscopic anisotropy and the presence of a nematic potential. The solution of the rotational diffusion equation for a rigid dipolar molecule in the presence of a nematic potential predicts that the frequency dependence of each component of the electric permittivity can be characterized by two exponential decays [25,26]. The rotation of a molecule around a short molecular axis in the presence of the nematic potential led to the low frequency relaxation mode **m₁** that is detected in the parallel component of the permittivity, which is classified as $\tau_{00}$ in terms of the spherical harmonics. The relaxation of the longitudinal component of the molecular dipole by precession around the director axis contributes to the high frequency mode **m₂** that is detected in the perpendicular component of the permittivity and is classified as $\tau_{10}$ in terms of the spherical harmonics.

For a nematic potential, it is assumed to be of the form:

$$U/k_B T = \sigma \cos^2 \theta \qquad (5)$$

where: $\sigma$ is proportional to the uniaxial order parameter $S$ (for $S<0.6$), while the approximate expressions for the relaxation times can be derived [26] as:

$$\sigma = \frac{3}{2} S(5 - \pi S)/(1 - S^2) \qquad (6)$$

The nematic potential barrier parameter, $\sigma$, is defined as $\sigma = q/RT$ where $q$ is height of the barrier that separates the two minima along the **n** direction.

$$\frac{\tau_{00}}{\tau_D} = \left(\frac{e^\sigma - 1}{\sigma}\right) \left(\frac{2\sigma\sqrt{\delta/\pi}}{1+\sigma} + 2^{-\sigma}\right)^{-1} \qquad (7)$$

$$\frac{\tau_{10}}{\tau_D} = \frac{1-S}{1+S/2} \tag{8}$$

The relaxation time, $\tau_D$, is the one for the rotational diffusion in the isotropic phase. The longitudinal, $\mu_l$, and transverse, $\mu_t$, components of the molecular dipole moment, $\mu$, contribute to the dielectric permittivity differently and they relax at different frequencies of the probe field. In solving the rotational diffusion equation, a uniaxial nematic potential was assumed. This conceals a basic inconsistency since the molecules that have a dipole that is inclined to the long molecular axis are intrinsically biaxial.

2. **Molecular modes in the nematic phase**

The dynamic dielectric responses of CBSCnSCB and CBSCnOCB were measured in metal cells that had a 50 μm spacer in a frequency range of 5Hz to 1 GHz. The dielectric response of CBSCnSCB and CBSCnOCB in the nematic phase, exhibited two relaxation processes, whose contribution to the dielectric spectrum depended on the orientation of the director in the measurement cell. According to the theoretical model of dielectric relaxation in nematic dimers [25,26], these dielectric relaxations could have been related to the rotational diffusion of the dimer molecules: the reorientation of the end-to-end dipolar groups parallel to the director at low frequencies (**m₁**), and the precessional movement of dipolar groups around the director for the high-frequency branch of the spectrum (**m₂**).

As was already mentioned, at higher frequencies, the $\varepsilon''$ spectra were dominated by two maxima. These corresponded to the two molecular relaxation modes **m₁** and **m₂** of the symmetric CBSCnSCB and asymmetric CBSCnOCB dimers, which are also made for CBnCB by Cestari *et al.* [5], Merkel *et al.* [24] and López *et al.* [40].

The two observed processes were simultaneously analyzed in the N phase, which enabled them to be described unequivocally by an orientational order, *S*, and the so-called Debye relaxation time, which corresponds to the relative relaxation in the absence of an orientational order. This can be interpreted simply as an extension of the isotropic relaxation time into the temperature range of the nematic phase. Following equation (4), the contribution to the perpendicular component of permittivity originated from the precession rotation of the transverse component of the dipole moment, i.e., the rotation of the bow axis of the bent-core conformation significantly contributed to the perpendicular component.

This was the assignment of mode **m₂**. The higher frequency relaxation process, m1, might have arisen from the rotations of a segment of the molecule, i.e., an internal rotation of each monomer with the spacer anchored that involved fluctuations of the cyanobiphenyl dipolar moment. Because the length of the spacer between the two mesogens in the dimer was large enough such an independent internal rotation of each monomer of the dimer was highly feasible. The temperature dependencies of the relaxation time $\tau_{10}$ are suggestive of a precessional rotation of the longitudinal component of each tio-cyanobiphenyl (SCB) dipole moment around the director in a planar-aligned cell.

The relaxation times of the modes, however, seemed to be well defined solely by their local orientation order in the nematic phase. For all of the dimers, the dynamics behavior in the nematic phase was reproduced quite well by the molecular dynamic model, Fig.1. The experimental relaxation times, $\tau_{00}$ and $\tau_{10}$, for modes **m₁** and **m₂**, respectively, were well fitted using eqs.(6-8) assuming $S$ and $\tau_D$ as unknown values. The results are presented in Fig 1, for $\tau_D$ and in insert for $S$. Although temperature dependencies of the orientational parameters corresponded to the results that were obtained by infrared spectroscopy quite well [29], considering the dielectric ones were related to the S-CB or (O-CB) dipole moments while the IR ones corresponded to the molecular axis.

In the $N_{TB}$ phase, the temperature dependencies of the $\tau_{10}$ maintained their trends without any step at the transition temperature despite the fact that the director was tilted with respect to the symmetry axis. It seemed that this process was exclusively determined by the local orientational order (relative to the local director), which continuously grew with a decreasing temperature. This was not the case for the end-over end relaxation time, $\tau_{00}$. First of all, the relaxation time dependence clearly showed a kink at the transition temperature, which indicated an increased relaxation rate at the transition temperature. This was more likely due to the critical fluctuations of the director in the vicinity of the transition [41]. Then, in the $N_{TB}$ phase, the temperature trend of the $\tau_{00}$ relaxation time suddenly increased with respect to the trend in the nematic phase. We simply interpreted this fact as the apparent growth of the potential barrier, $q$, that separated the two minima along the *n* direction after entering the $N_{TB}$ phase. This finding corresponded well with the orientational correlations effect of the longitudinal dipoles ($g_{\parallel} \neq 1$) that was revealed by IR spectroscopy [28,29].

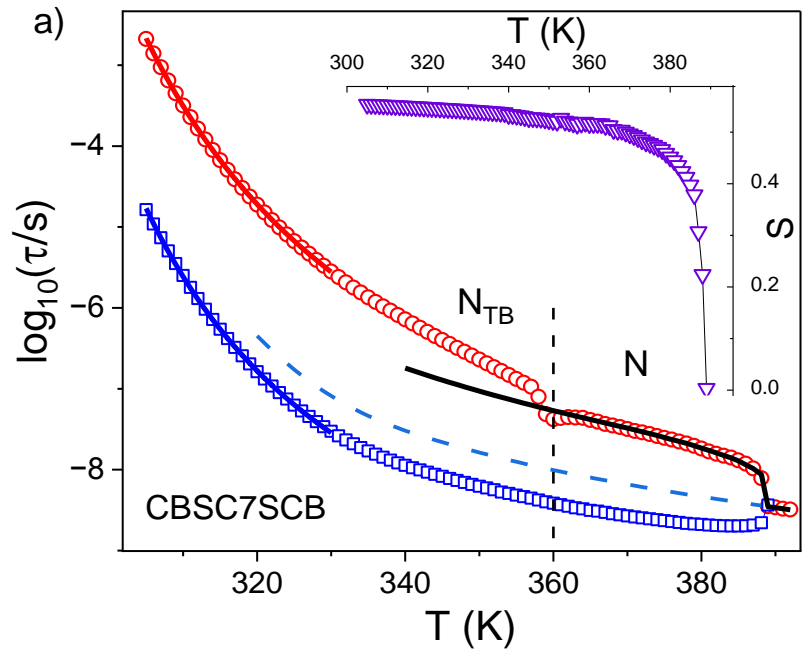

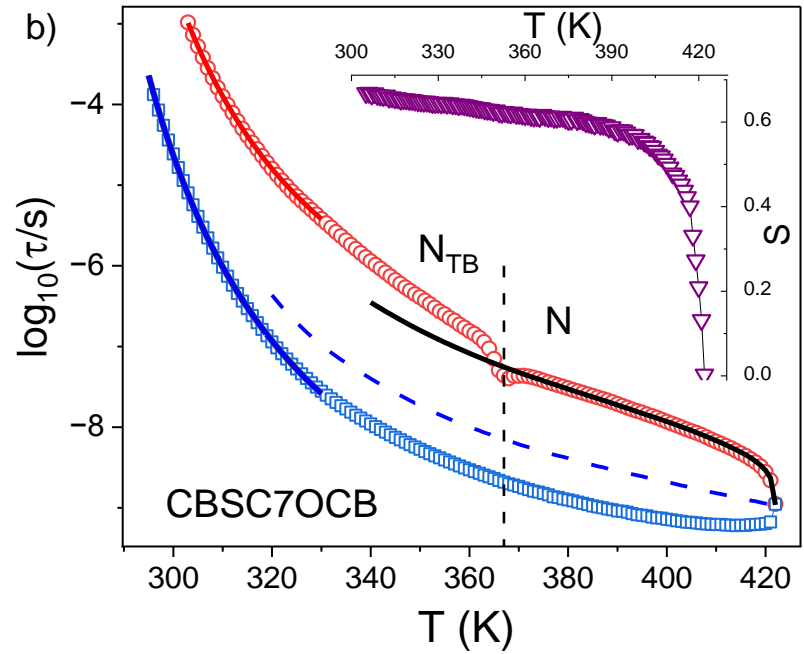

**Fig. 1.** Plots of the relaxation times for modes $m_1$ and $m_2$ (5μm planar cell): ☐ - $m_1$ mode, ○ - $m_2$ mode, model fitting : — $m_1$ mode, -- $\tau_D$ relaxation time and ▽ - order parameter, S (as an insert) (a) – CBSC7SCB dimer, (b) – CBSC7OCB.

Although the models that are used to interpret the low-frequency relaxation in liquid crystals are often based on a single particle relaxation process, spectroscopic probes of the molecular motion such as magnetic resonance, neutron scattering and time-resolved fluorescence depolariastion, suggest that reorientation times for mesogens are of the order of $10^{-9}$ s to $10^{-10}$ s in the isotropic, nematic and disordered smectic phases. Thus, interpreting the dielectric relaxation processes at MHz or even kHz frequencies in terms of the rotation of a single molecule is not likely to be correct. The low-frequency relaxations that were observed in liquid crystals were the result of a collective molecular motion, although the models outlined above are useful for analyzing the results and comparing materials.

## 3. Dynamic characterization of glass formation

We will now focus on the temperature range below the transition N-$N_{TB}$. Only two longer dimers with seven carbons in the link, i.e., CBC7SCB and CBSC7OCB exhibited the typical glass-forming behavior after cooling. The phenomenological equations, the Vogel-Fulcher-Tammann formula eq. (2) were required to describe the temperature dependence of the relaxation time data ($\tau$). The results of the fitting are shown as dashed lines for $\tau_D$ in and open triangles for *S*-parameter in Figures 1a and 1b.

The VFT parameters that were obtained for eq.(2) are listed in Table I. It should be stressed that as can be listed in Table I that two different dielectric glass transition temperatures were obtained though they were rather close to one another within about 1 K. It was also interesting to note that the pre-factor $\tau_0$ for the **m₂** mode was of the order of $10^{-11}$ s.

TABLE I. Fitting parameters according to Eq. (2) for the different dimers and the calculated glass transition temperature for the **m₁** and **m₂** modes

|      | mode  | $\log_{10}[\tau(s)]$ | $T_0$ (K) | B (K) | $T_{gl}$ (K) |
|------|-------|----------------------|-----------|-------|--------------|
| CS7S | $m_1$ | -10.1                | 266.6     | 682.6 | 289.8        |
|      | $m_2$ | -11.9                | 265.6     | 631.9 | 286.3        |
| CS7O | $m_1$ | -9.2                 | 261.3     | 596.0 | 284.3        |
|      | $m_2$ | -11.5                | 260.5     | 621.7 | 280.5        |

A more detailed method for studying the dynamics of glass forming behavior is to analyze the derivatives of the relaxation time with respect to the inverted temperature [42,43]. The application of this procedure by Rzoska and Drozd-Rzoska led to the relationship:

$$\left[\frac{d\ln\tau}{d(1/T)}\right]^{-1/2} = \left[\frac{H_A(T)}{R}\right]^{-1/2} = B^{-1/2}(1 - T_0/T), \qquad (9)$$

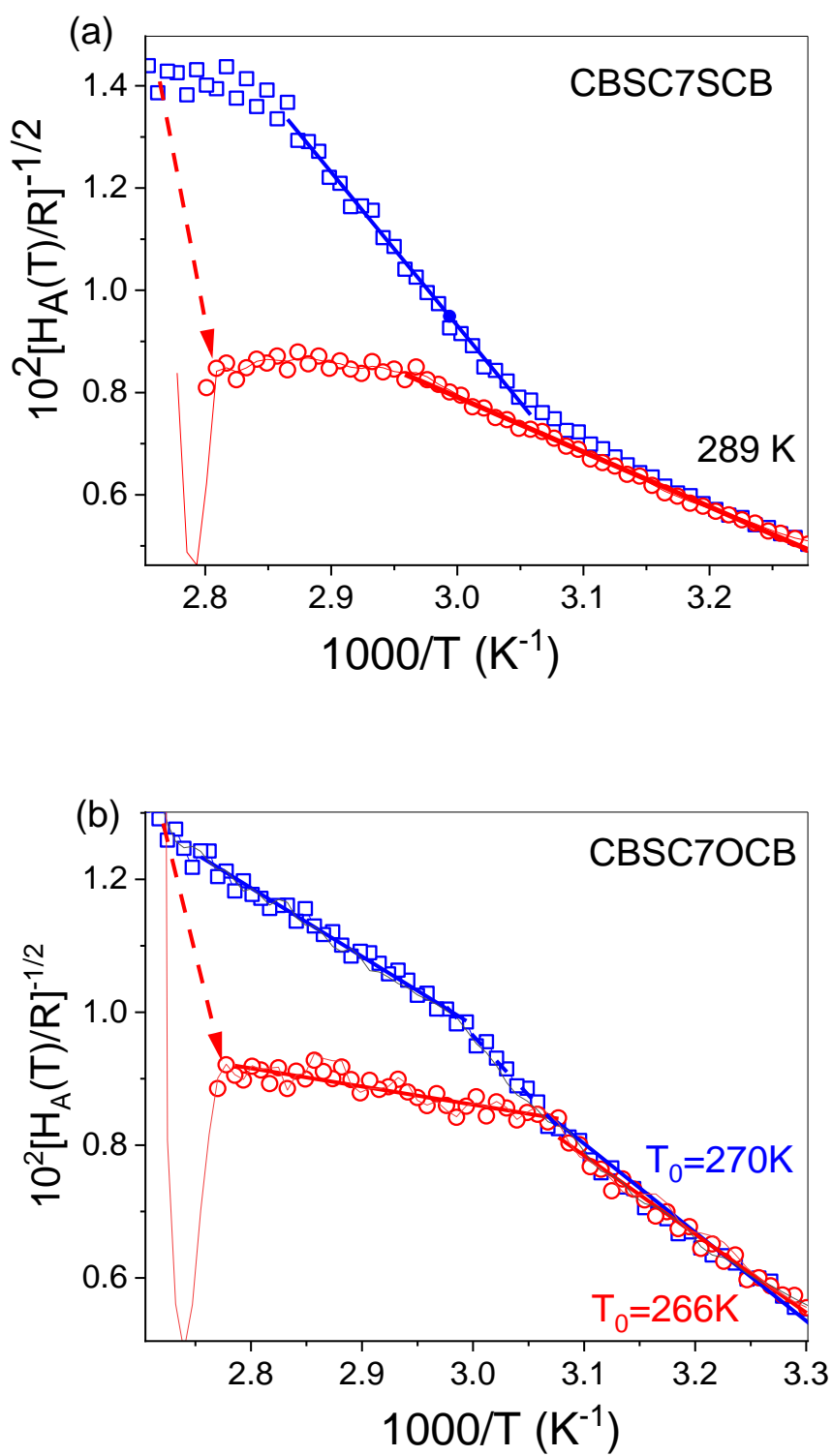

**Fig. 2.** Plots show a linear dependence $[H_A(T)/R]^{-1/2}$ on the inverse temperature (a) – CBSC7SCB dimer, (b) – CBSC7OCB. The arrows show the jump of the $H_A$ that omits the critical fluctuation at the N-N$_{TB}$ transition, where $H_A$ is denoted as the apparent enthalpy of the activation. □ - **m₁** mode, ○ - **m₂** mode.

In terms of the validity of the VFT equation, Eq. (2) was predicted to have a linear dependence $[H_A(T)/R]^{-1/2}$ on the inverse temperature. The results of applying this equation to both modes are shown in Fig.2a for CBSC7SCB and Fig 2b for CBSC7OCB. In the case of CBSC7SCB and mode $m_2$, the activation enthalpy (almost constant in N phase $H_A$=42 kJ/mol) began to increase in the $N_{TB}$ phase and exhibited a region of linearity (of $[H_A(T)/R]^{-1/2}$) down to 325 K (1000/T≅3.08 K$^{-1}$). Conversely, the activation enthalpy of the mode $m_1$ jumped at transition N-$N_{TB}$ (from 54 kJ/mol to 113 kJ/mol) then increased, although less rapidly, to join $m_2$ below 325 K and from this temperature both modes had the same enthalpy of activation, probably due to their cooperative behavior. Similarly, the mode $m_2$ for CBSC7OCB gradually increased the activation enthalpy in the $N_{TB}$ phase (from 53 kJ/mol at transition temperature up to 116 kJ/mol and also exhibited a linear region ($[H_A(T)/R]^{-1/2}$) down to 330 K (1000/T≅3.03 K$^{-1}$). In this case, the mode $m_2$ had two linear domains: the first from the N-$N_{TB}$ transition down to 330 K and the second below 330 K. This coincided with the temperature at which the lateral/transversal interaction began to be important [29].

TABLE II Fitting parameters according to Eq. (2) & Eq. (9) for the different dimers.

|  | $\log_{10}[\tau(s)]$ | Eq.(2) | Eq.(9) | range [1000/T(K$^{-1}$)] |
|---|---|---|---|---|
| CBSC7SCB | -10.3 | 269 | 289 | 2.96-3.3 |
|  | -11.6 | 264 | 289 | 3.15-3.3 |
| CBSC7OCB | -8.9 | 265 | 266 | 3.08-3.3 |
|  | -11.5 | 261 | 270 | 3.08-3.3 |

The activation enthalpy of the mode $m_1$ jumped again at the transition N-$N_{TB}$ (from 53 kJ/mol to 101 kJ/mol) and then grew, but less rapidly, to join the $m_2$ mode at about 320 K after which both modes had the same activation enthalpy (~116 kJ/mol) due to their cooperative behavior. Dashed and continuous lines in both figures indicate the linear fits according to eq.(6-8).

We now consider the temperature-derivative procedure that should be used for a critical-like description through Eqs. (3). A linear dependence with temperature was obtained in the region of the validity of the critical-like descriptions according to the relationship:

$$T^2\left[\frac{d\ln\tau}{d(1/T)}\right]^{-1} = \left[\frac{H_A(T)}{T^2 R}\right]^{-1} = \Phi^{-1}(T - T_c), \tag{10}$$

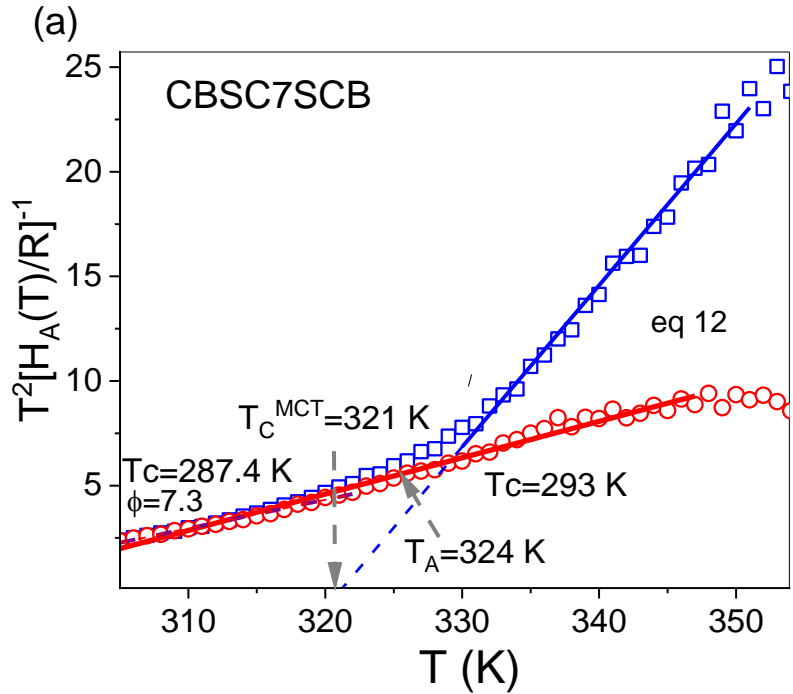

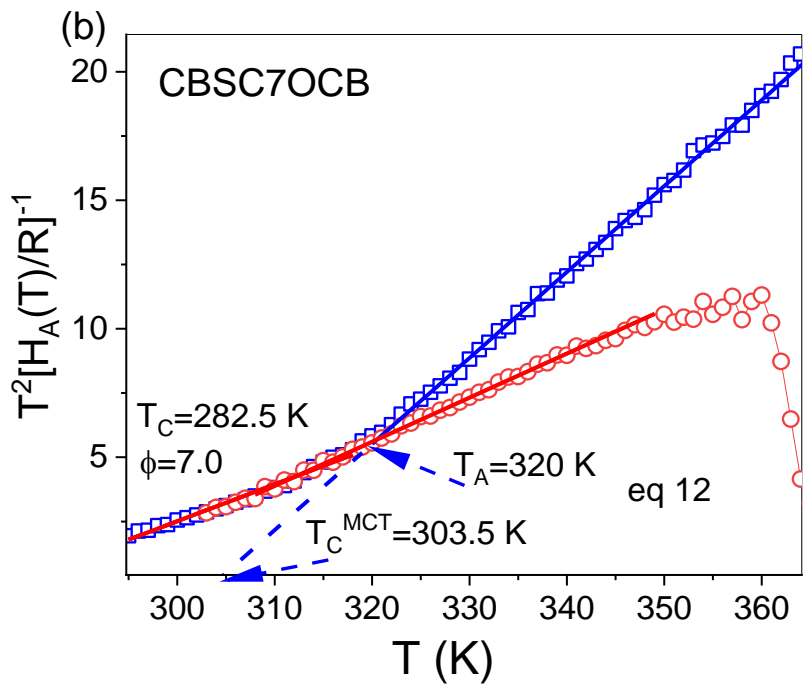

**Fig 3.** Results of the temperature-derivative analysis [eq. (10)] that was applied to both the **m₁** and **m₂** modes (☐ - **m₁** mode, ○ - **m₂** mode) of (a) CBSC7SCB and (b) CBSC7OCB in which the linear dependencies indicated the domains of the validity of the critical-like description. Solid and dashed lines correspond to the dynamic domains for the **m₁** and **m₂** modes, respectively, according to Eq. (10). The fitting parameters are listed in Table III.

The $T_C$ in the above MCT critical-like equation accounts for the crossover temperature from the ergodic to the non-ergodic domain. The critical temperature $T_c$ seemed to correlate with the caging temperature $T_A$. The experimental data for both modes are shown in Figure 3 as a plot of $[T^2R/H_A(T)]$ vs. temperature. It should be stressed that for both dimers, the data of the m1 mode exhibited a linear behavior, i.e., followed the DS-model, over the entire temperature range of the $N_{TB}$-mesophase. The temperature $T_C$ was the temperature of the virtual phase transition (the critical temperature), which was usually located slightly below $T_g$,. The exponent $\Phi \approx 6\text{-}15$ was suggested as being universal for the glass-forming polymers. In contrast, the **m₂** data in the $N_{TB}$-mesophase clearly exhibited two linear domains, one at low temperatures (the DS-model) from $Tg$ up to about 320 -324 K (denoted as $T_A$ in the figures); they seemed to be indistinguishable from the behavior of the **m₁** mode. and another domain at high temperatures (MCT-like description) from $T_A$ up to the vicinity of the $N_{TB}$-N phase transition for both modes. The cross-over temperature coincided well with the IR spectroscopy observation for the temperature when the lateral interactions increased their strength [29]. Below that temperature, it seemed that both modes became joined and were described in the same way.

The subsequent linear fittings according to Eq. (9) yielded the values of the parameters for each mode and dynamic domain, namely $T_C$ and the exponent $\Phi$. The final fitting of the relaxation data according to Eq. (9) was focused on the pre-factor $\tau_0$. All of the derived parameters are listed in Table III and the results of the fittings are presented in Figure 3.

TABLE III. Fitting parameters according to Eq. (10) for the different dimers.

|  | $T_A$ | $T<T_A$ | $T_c$ | $\Phi$ | range(K) | description |
|---|---|---|---|---|---|---|
| CBSC7SCB | 324 K | 287.4 | 293 | 7.3 | 305-347 | DS |
|  |  | 7.3 | 321 | 1.3 | 330-351 | MCT |
| CBSC7OCB | 320 K | 282.5 | 287.4 | 7.0 | 295-350 | DS |
|  |  | 7.0 | 303.5 | 1.3 | 320-365 | MCT |

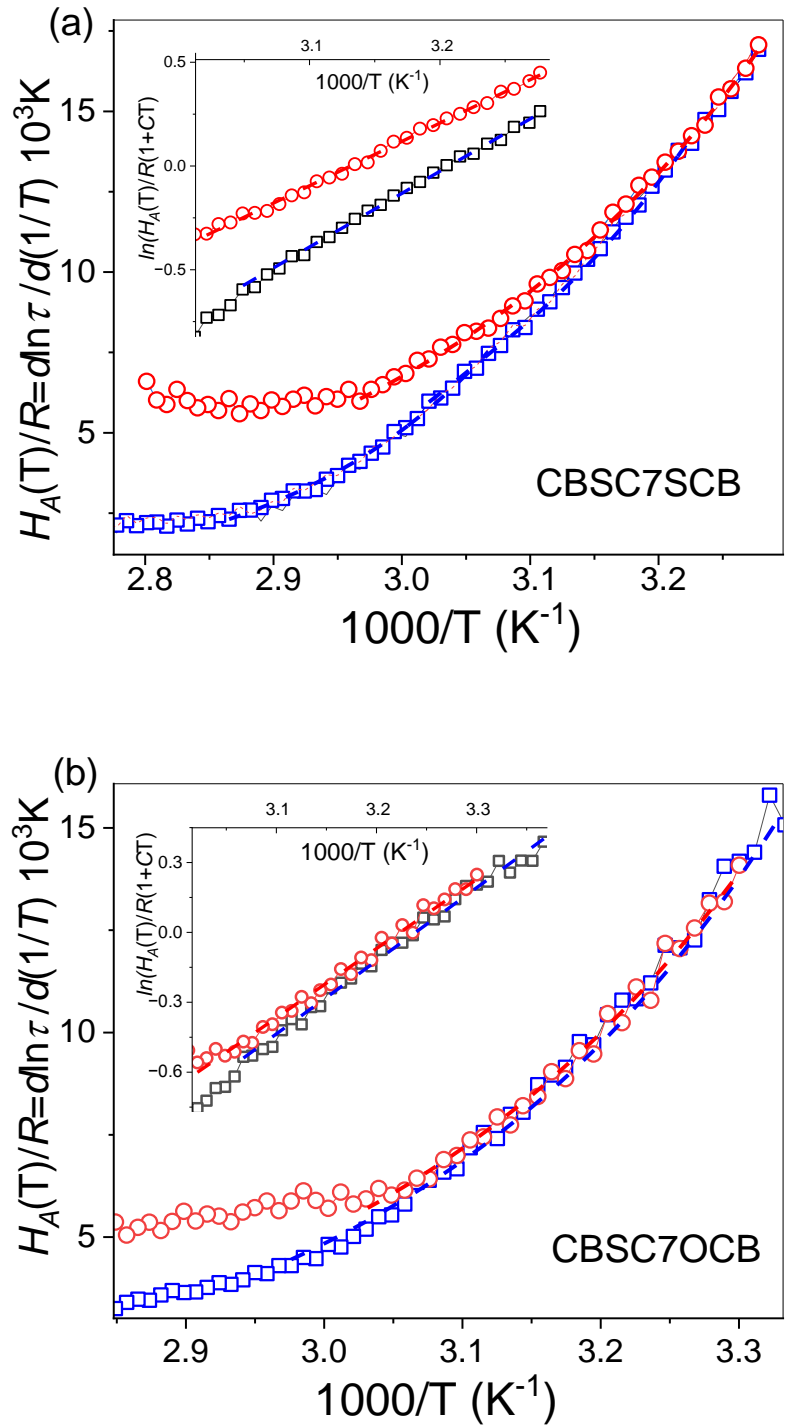

Fig 4. Results of the temperature-derivative analysis [eq. (11)] that was applied to both the **m₁** and **m₂** modes of (a) CBSC7SCB and (b) CBSC7OCB that (solid and dashed lines) indicate the different dynamic domains of the validity of the critical-like description (of eq.10). □ - **m₁** mode, ○ - **m₂** mode.

One of the most noticeable results in Table III was related to both the $T_C$ and $T_g$ temperatures that corresponded to the **m₁** and **m₂** modes. It seemed that a common critical temperature ($T_C \approx 290$ K for CBSC7SCB and $T_C \approx 287$ K for CBSC7OCB) for the virtual phase transition as well as a same glass transition temperature was obtained for both modes. As for the exponent, in the DS-domain, $\Phi$ is 7.0 and 7.3 (for **m₁**-mode) are the typical values that are usually reported for spin-glass-like systems. In the MCT-domain for the m₂-mode, $\Phi$ was within the usual range of values.

The application of the temperature-derivative procedure to Eq. (4) did not permit a similar simple linearization methodology as the previous eq. Instead, an enthalpy function was described by two uncorrelated variables ($K$ and $C$) in the form:

$$\left[\frac{d\ln\tau}{d(1/T)}\right] = \left[\frac{H_A(T)}{R}\right] = K(1 + C/T)e^{C/T} \qquad (11)$$

Table IV. Fitting parameters according to Eq. (11) for the different dimers.

|  | $\log_{10}[\tau(s)]$ | K(K) | C(K) | $T_{gl}$ (K) | range [1000/T(K⁻¹)] |
|---|---|---|---|---|---|
| CBSC7SCB | -7.78 | 0.0789 | 3016 | 268.2 | 2.97-3.28 |
|  | -9.08 | 0.0082 | 3653 | 267.9 | 3.02-3.22 |
| CBSC7OCB | -7.15 | 0.058 | 3028 | 263.9 | 3.03-3.30 |
|  | -9.05 | 0.031 | 3198 | 260.6 | 2.98-3.35 |

The fitting procedure was unable to fit to only one dynamic domain for each mode as can be observed in Figure 4. however, it enabled us to better distinguish the different dynamic domains. It should be stressed that according to Eq. (11) and the results of the fits for the $C$ and $K$ parameters (see Table IV), $\ln[H_A(T)/R]$ exhibited a nearly linear dependence with the inverse temperature. As was suggested recently [43], to proceed further with a successful linearized expression, Eq. (11) can be written as:

$$\ln\left[\frac{H_A(T)}{R(1+C/T)}\right] = \ln K + C/T \qquad (12)$$

The final fitting of the relaxation time data according to Eq. (4) led to the pre-factor $\tau_0$. All of the (WM)-fitting parameters $C$, $K$, and $\tau_0$, for each dynamic domain and relaxation mode, are listed in Table IV. The final fits are also presented in Figure 4. The results for $\tau_0$ and $T_g$ (see Table IV) indicated a strong parallelism with the VFT-results. Again, $\tau_0$ for the m₂-mode was of the order of $10^{-10}$ s (high temperature dynamic domain), which was far from $10^{-14}$ s. As for the glass transition temperature, two different values were obtained, ca. but less than 3K apart.

**CONCLUSIONS**

The dielectric relaxation spectra for the symmetric CBSC7SCB and asymmetric CBSC7OCB dimers was resolved in terms of two relaxation processes: **m₁** and **m₂**, as was predicted by the rotational diffusion model. The high-frequency relaxation process originated from the precession rotation of the longitudinal component of the dipole moment of the monomers of each thioether cyanobiphenyl (SCB) dipole moment around the director. The relaxation rate accelerated with respect to the isotropic relaxation because $\tau_{10}$ was shorter than $\tau_D$. The low-frequency relaxation corresponded to originated from the "end-over-end" rotation of the dipole moment of the monomers. The relaxation rate of that mode was lower with respect to the isotropic one. It was clear that the dynamic of the two relaxation processes modes, **m₁** and **m₂**, were quite different in the two temperature regions of the $N_{TB}$ mesophase. In the first region, which extended almost 30 K from the N-$N_{TB}$ transition, the mode **m₁** assigned it to an "end-over-end" rotation, the activation enthalpy initially jumped almost twice on entering $N_{TB}$ and then increased continuously as the temperature decrease in range of almost 30 K. This sudden increase in enthalpy was found to be associated with the appearance of the longitudinal dipole correlations that were revealed in the IR studies. In the same region, the activation enthalpy of mode **m₂** (assigned to the precession of the dipoles) grew continuously but much faster than **m₁** and finally, they met at the same value of $H_A(T)$. This meeting temperature corresponded well to the appearance of the transverse interactions as was observed in the IR measurements.

The description of the critical behavior of the dynamic domains that are related to the cooperative molecular motions enough to adequately describe the dynamics that were associated with the two main molecular motions in the $N_{TB}$ mesophase of the symmetric and asymmetric dimers. Both types of molecular motion appeared to strongly cooperate at low temperatures, and changed in a coordinated manner as the temperature of the material approached the glass transition point. As was expected, it was determined that both of the molecular motions determined the glassy dynamics at the same glass transition temperature, $T_g$.

**Acknowledgments**: Authors K.M. and A.K. thank the National Science Centre for funding through grant no. 2018/31/B/ST3/03609. All DFT calculations were carried out with the

Gaussian09 program using the PL-Grid Infrastructure on the ZEUS and Prometheus clusters in Academic Computer Center CYFRONET AGH (AGH University of Science and Technology) in Cracow, Poland.